\documentstyle[12pt,fleqn]{aipproc}
\renewcommand{\baselinestretch}{1.0}

\def\beeq{\begin{equation}}
\def\eneq{\end{equation}}
\def\beeqa{\begin{eqnarray}}
\def\eneqa{\end{eqnarray}}
\def\theequation{\arabic{equation}}

\def\thesection{\arabic{section}.}
\def\thesubsection{\arabic{section}.\arabic{subsection}}
\def\thepage{-- \arabic{page} --}
\addtocounter{section}{-1}
\begin{document}

\begin{center}

\mbox{}

\mbox{}

{\Large {\bf Mechanism of Magnetism in Stacked\\
Nanographite: Theoretical Study}}

\mbox{}

\mbox{}

{\large Kikuo Harigaya$^{* \dagger}$, 
Naoki Kawatsu$^\dagger$, and Toshiaki Enoki$^\dagger$}

\mbox{}

\mbox{}

{\small {\sl $^*$Electrotechnical Laboratory, 
Umezono 1-1-4, Tsukuba 305-8568, Japan\\
$^\dagger$Tokyo Institute of Technology,
Ookayama 2-12-1, Meguro-ku 152-8551, Japan}}

\end{center}

\begin{abstract}
Antiferromagnetism in stacked nanographite is investigated 
with using the Hubbard-type model.  The A-B stacking is favorable
for the hexagonal nanographite with zigzag edges, in order 
that magnetism appears.  Next, we find that the open shell 
electronic structures can be origins of the decreasing 
magnetic moment with the decrease of the inter-graphene
distance, as experiments on adsorption of molecules suggest.
\end{abstract}

\mbox{}

\begin{center}
{\large {\bf INTRODUCTION}}
\end{center}

\mbox{}

Nanographite systems, where graphene sheets of the orders of 
the nanometer size are stacked, show novel magnetic properties, 
such as, spin-glass like behaviors [1], and the change of 
ESR line widths while gas adsorptions [2]. Recently, it 
has been found [3] that magnetic moments decrease with the 
decrease of the interlayer distance while water molecules 
are attached physically.

In this paper, we consider the stacking effects 
in order to investigate mechanisms of antiferromagnetism 
using the Hubbard-type model with the interlayer hopping
$t_1$ and the onsite repulsion $U$.  We will show that 
the A-B stacking is favorable for the hexagonal nanographite 
with zigzag edges, in order that magnetism appears.
Next, we show that the open shell electronic structures, 
coming from functional units and/or geometrical effects, 
can be origins of the decreasing magnetic moment with 
adsorption of molecules.  Details will be reported elsewhere [4].

\mbox{}

\begin{center}
{\large {\bf CLOSED SHELL ELECTRON SYSTEMS}}
\end{center}

\mbox{}

First, we report the total magnetic moment per layer 
for the A-B stacked hexagonal nanographite shown in Fig. 1 (a).
The first and second layers are displayed by the thick and
thin lines, respectively.  In each layer, the nearest neighbor
hopping $t$ is considered. Each layer has closed shell electron
systems when the layers do not interact mutually, because the 
number of electrons is equal to the number of sites.  The 
interlayer hopping $t_1$ is assigned at the sites 
with closed circles.  The model is solved with the unrestricted
Hartree-Fock approximation, and antiferromagnetic solutions
are obtained.  Figure 1 (b) shows the absolute value of 
the total magnetic moment per layer as functions of $t_1$ and $U$.
As increasing $U$, the magnitude of the magnetization increases.  
The magnetic moment is zero at the smaller $t_1$ region 
for $1.9t$ (open squares), $2.0t$ (closed circles), and $2.1t$
(open circles).  The magnetic moment is zero only at $t_1 = 0$
for $U=2.2t$ (closed triangles) and $2.3t$ (open triangles).  
We can understand the parabolic curves as a change due to the 
Heisenberg coupling proportional to $t_1^2/U$.

\begin{figure}[tbp]
\vspace*{5.3cm}
\caption{(a) A-B stacked hexagonal nanographite with 
zigzag edges. (b) The absolute magnitude of the total magnetic
moment per layer as a function of $t_1$.  The onsite interaction
is varied within $1.8t$ (closed squares) $\leq U \leq 2.3t$ 
(open triangles).  The interval of $U$ between the series of 
the plots is $\Delta U = 0.1t$.}
\end{figure}%

We have also calculated for the simple A-A stacking.  
We have not found any finite magnetization in this case.  
This is a remarkable difference between the A-A and A-B 
stackings, and is a new finding of this paper.  
The A-B stacking should exist in nanographite systems,
because the exotic magnetisms have been observed
in recent experiments [1-3].  The decrease of the 
interlayer distance while attachment of water molecules 
makes $t_1$ larger.  However, it is known that the 
magnetism decreases while the attachment of molecules [3].
The calculation for the closed electron systems
cannot explain the experiments even qualitatively.

\mbox{}

\begin{center}
{\large {\bf OPEN SHELL ELECTRON SYSTEMS}}
\end{center}

\mbox{}

Here, we consider the Hubbard-type model for systems which 
have open shell electronic structures when a nanographene 
layer is isolated.  One case is the effects of additional 
charges coming from functional side groups.  The next case 
is the roles of the standing magnetic moments due to the 
geometrical origin.

\begin{figure}[tbp]
\vspace*{5.3cm}
\caption{The absolute magnitude of the total magnetic
moment per layer as a function of $t_1$ for the system with
a site potential $E_s = -2 t$, (a) at the site A and (b) at 
the site B.  The site positions are displayed in Fig. 1 (a).  
In (a), the onsite interaction is varied within $0.6t$
(closed squares) $\leq U \leq 1.8t$ (closed triangles).
The interval of $U$ between the series of the plots
is $\Delta U = 0.3t$.  In (b), it is varied within $1.0t$ 
(closed squares) $\leq U \leq 2.0t$ (closed triangles).
The interval of $U$ between the series of the plots
is $\Delta U = 0.25t$.}
\end{figure}%

The active functional groups are simulated with introducing
a site potential $E_s$ [5] at edge sites.  When $E_s > 0$,
the site potential means the electron attractive groups.
When $E_s < 0$, the electron donative groups are simulated
because of the increase of the electron number at the
site potentials.  Here, we take $E_s = -2t$, and one
additional electron per layer is taken account.
Figure 2 displays the absolute values of total magnetic
moment per layer.  In Fig. 2 (a), the site potentials 
locate at the site A in the first layer [Fig. 1 (a)], 
and at the symmetry equivalent site in the second layer.  
In Fig. 2 (b), the site potential exists at the site B.
The total magnetization is a decreasing function
in both figures.  The decrease is faster in Fig. 2 (b)
than in Fig. 2 (a).  The site B is neighboring to the
site with the interaction $t_1$, and thus the localized
character of the magnetic moment can be affected easily
in this case.  The decease of magnetization by the
magnitude $30-40$\% with the water molecule attachment [3]
may correspond to the case of Fig. 2 (b).

\begin{figure}[tbp]
\vspace*{5.3cm}
\caption{(a) A-B stacked triangulene with vertical shift.
(b) The absolute magnitude of the total magnetic moment per layer 
as a function of $t_1$.  The onsite interaction
is varied within $0.4t$ (closed squares) $\leq U \leq 2.0t$
(closed triangles).  The interval of $U$ between the series of the plots
is $\Delta U = 0.4t$.}
\end{figure}%

Next, we look at the magnetism of stacked ``triangulenes".
The ``triangulene" has the geometry displayed in Fig. 3 (a),
and there are nine hexagonal rings [6].  The Lieb's theorem 
[7] says that the total spin $S_{\rm tot}$ of the repulsive 
Hubbard model of the A-B bipartite lattice is 
$S_{\rm tot}=\frac{1}{2}| N_A - N_B |$, where $N_A$ and $N_B$ 
are the numbers of A and B sites.  We find $S_{\rm tot} = 1$
for the single triangulene.  Figure 3 (b) displays the 
absolute magnitude of the total magnetic moment per layer 
for the A-B stacking with the vertical shift [Fig. 3 (a)]. 
The total magnetic moment is a decreasing function with 
respect to $t_1$.  As we discuss in detail [4], there appear 
strong local magnetic moments at the zigzag edge sites, 
and they give rise dominant contributions to the magnetism 
of each layer.  In the triangulene case, most of the edge
sites are neighboring to the sites with the interaction $t_1$
in Fig. 3 (a).  The interactions of the edge sites with the 
neighboring layers are strong, and the itinerant characters 
of electrons become larger as increasing $t_1$.  Therefore, 
the magnetic moment is a decreasing function in Fig. 3 (b).

The present two calculations agree with the experiments, 
qualitatively.  We can explain the decrease of magnetism
in the process of adsorption of molecules [3].
Thus, the open shell electronic structures due to 
the active side groups and/or the geometrical origin 
are candidates which could explain the exotic magnetisms.

\mbox{}

\begin{center}
{\large {\bf SUMMARY}}
\end{center}

\mbox{}

Antiferromagnetism in stacked nanographite has been 
investigated with the Hubbard-type model.  The A-B 
stacking is favorable for the hexagonal nanographite 
with zigzag edges, in order that magnetism appears.  
Next, we have found that the open shell electronic 
structures can be origins of the decreasing magnetic 
moment with adsorption of molecules.

\mbox{}

\begin{center}
{\large {\bf REFERENCES}}
\end{center}

\mbox{}

\noindent
1. Y. Shibayama {\sl et al.}, 
Phys. Rev. Lett. {\bf 84}, 1744 (2000).\\
2. N. Kobayashi {\sl et al.},
J. Chem. Phys. {\bf 109}, 1983 (1998).\\
3. N. Kawatsu {\sl et al.},
Meeting Abstracts of the Physical Society of Japan
{\bf 55} Issue 1, 717 (2000).\\
4. K. Harigaya, J. Phys.: Condens. Matter {\bf 13}, 
in press (2001); cond-mat/0010043; cond-mat/0012349.\\
5. K. Harigaya, A. Terai, Y. Wada, and K. Fesser,
Phys. Rev. B {\bf 43}, 4141 (1991).\\
6. G. Allinson, R. J. Bushby, and J. L. Paillaud,
J. Am. Chem. Soc. {\bf 115}, 2062 (1993).\\
7. E. H. Lieb, Phys. Rev. Lett. {\bf 62},
1201 (1989); {\sl ibid.} {\bf 62}, 1927 (1989).\\

\end{document}